\documentclass[aps,prb,preprint,superscriptaddress,floatfix,showpacs]{revtex4}

\usepackage{graphicx}

\usepackage[intlimits]{amsmath}
\usepackage{amscd}
\usepackage{amssymb}
\usepackage{amsfonts}
\usepackage{amsthm}

\newcommand{\noshow}[1]{}
\newcommand{\myfigref}[1]{\mbox{Fig.\,\ref{#1}}}

\begin{document}


\title{Transmission in the vicinity of the Dirac point in hexagonal Photonic Crystals}


\author{Marcus Diem}
\email[]{diem@ameslab.gov}
\affiliation{Ames Laboratory and Department of Physics and Astronomy, Iowa State University, Ames, IA, 50011}
\author{Thomas Koschny}
\author{C. M. Soukoulis}
\affiliation{Ames Laboratory and Department of Physics and Astronomy, Iowa State University, Ames, IA, 50011}
\affiliation{Institute of Electronic Structure (IESL) and Laser, Foundation for Research Technology Hellas (FORTH) and
Department of Material Science and Technology, University of Crete, 71110 Heraklion, Crete, Greece}


\date{\today}

\begin{abstract}
We use a scattering matrix approach to simulate the transmission through a hexagonal
Photonic Crystal in the vicinity of the Dirac point. 
If the crystal is oriented so that the propagation direction perpendicular
to the surface corresponds to the $\Gamma$K direction, no oblique transmission is
possible for a very long (infinite) structure. 
For a finite structure with width, W, and length, L, the length dependence of the transmission is given
by $\mathrm{T}_\mathrm{total}=\Gamma_0\mathrm{W}/\mathrm{L}$. For $\mathrm{T}_\mathrm{total}$ all waves with
a wavevector parallel to the surface, $k_\shortparallel=n\frac{2\pi}{\mathrm{W}}$, described by a
channel number, $n$,  must be considered. 
We show the transmission at the Dirac point follows the given scaling law and this scaling
law is related to the behavior of the individual channels. This leads to the establishment
of a criterion for the maximum length for
this scaling behavior when the total transmission reaches a constant value. We also compare this
scaling behavior to the results in other frequency regions.
\end{abstract}

\pacs{42.70.Qs,41.20.Jb,42.25.Bs }


\maketitle




%

\section{\label{sec:intro}Introduction}
The simulation of two-dimensional Photonic Crystals (PC) with a hexagonal lattice
has so far primarily focused on studying the band gap, either to obtain the largest possible
gap\cite{oe.12.5684,PhysRevB.52.R2217,PhysRevB.53.7134,PhysRevB.44.8565,cassagne:289,gadot:1780}
or to study the impact of disorder on the width of the
gap\cite{LighLocalSou01,PhysRevB.59.5463,PhysRevB.66.113101,PhysRevB.61.13458}
or wave-guiding properties in such crystals.\cite{Gerace:04}  The transmission
in the band regions was used to characterize experimental samples\cite{VKMWDGPBSWG03} or for
studies of negative refraction.\cite{PhysRevB.62.10696,FS03,PhysRevLett.93.073902,PhysRevB.70.205125,PhysRevLett.92.127401,moussa:085106,gajic:165310}
For a general overview see review by Busch et al. and references therein.\cite{PCreview}

Recently, Raghu and Haldane pointed out that the K-point in the band-structure can also be seen
as the optical analogue to the Dirac point in graphene.\cite{2006cond.mat.2501R,2005cond.mat.3588H}
 At these points the band-structure exhibits a conical
singularity with a linear dispersion relation, as it occurs in the Dirac equation.
This offers the possibility to discuss many interesting effects of this dispersion predicted
in the electronic case for graphene,\cite{natMatGraphene,KatGraphene} such as changes in the
conductance fluctuations\cite{0295-5075-79-5-57003} and enhanced transmittance in the
disordered case\cite{ostrovsky:256801,o:246802} in a non-interacting photonic system.

Around the Dirac point, a pseudo-diffusive transmission behavior characterized by
a scaling of the transmission proportional to W/L with the width, W, and length, L, of
the PC was predicted.\cite{sepkhanov:063813}
This result was obtained in an analytic approach by discretizing the incoming modes
into channels with a spacing of the wavevector parallel to the surface, $k_{\shortparallel}$, by $\Delta k_\shortparallel\,$=$\,\frac{2\pi}{\mathrm{W}}$
and by using current conservation and symmetry relation in a transfer matrix approach analog to a
calculation for graphene.\cite{Kat08,o:246802} Numerical studies, using the 
multiple-scattering Korringa-Kohn-Rostocker method\cite{zhang07graphene} or finite
difference time-domain\cite{SB07} confirmed the results.
However, in both numerical approaches, only short crystals and only a small number of lengths have been studied
and the behavior of individual channels has been ignored as well.

In this paper, we investigate the
contribution of these channels and show there is also a width-dependent upper length limit for the W/L scaling for
long crystals associated with the complete suppression of all channels except the $0^{\mathrm{th}}$ one with
$k_\shortparallel\,$=$\,0.0$. This behavior is very important in understanding the predicted enhancement of transmittance
at the Dirac point in disordered Photonic Crystals.\cite{sepkhanov:063813}

\begin{figure}
\includegraphics[width=7.5cm]{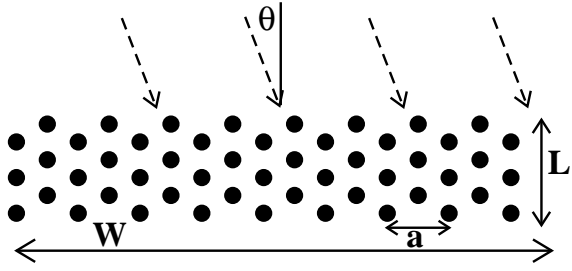}
\caption{\label{fig1}Structure. The shown example has a length of 3 unit cells and a width of 8. A
plane wave with an angle $\theta$\,=$\,0$ (perpendicular to the surface) propagates along the $\Gamma$K
direction in the band-structure. One unit cell in the propagation direction contains 2 rows of cylinders.}
\end{figure}

We use our own implementation of a  Fourier-Modal method with a scattering matrix approach, also known as
Rigorous-Coupled Wave Analysis (RCWA),\cite{MG81,WC99} which allows us to simulate crystals of arbitrary length, L,
and to determine the limit on $W>>L$ not discussed in previous publications. Special care is taken of the correct Fourier-factorization rules
to ensure a fast convergence.\cite{LiMath96,Li2D97,He02} This approach assumes incoming plane waves,
defined by a dimensionless frequency $\omega'\,$=$\,\omega a/2\pi c\,$=$\,\mathrm{a}/\lambda$ and
the angle $\theta$ to the surface normal (\myfigref{fig1})
onto a periodic structure with lattice constant $a$.

The transmittance, T, for one frequency-angle pair through the structure
is calculated by summing over all propagating diffraction orders and adding up the magnitude
of their Poynting vectors. The same holds for the reflectance, R, and the sum of both is
tested to be equal to unity. In the studied case
the finite width is incorporated by the superposition of 2$N$+1 planes waves with
different $k_\shortparallel$ corresponding to the channels
discussed above. Each channel with number $n$ is associated with a $k_{\shortparallel,n}\,$=$\,n\frac{2\pi}{\mathrm{W}}$ and an angle
to the surface normal given by $\theta\,$=$\,\mathrm{arcsin}(\frac{k_\shortparallel}{k_0})$ with $k_0\,$=$\,2\pi\omega'$.
The maximum/minimum parallel component of the wavevector is then given by $\pm N \frac{2\pi}{\mathrm{W}}$. For crystals with a length
$L>1/k_{\shortparallel,\mathrm{max}}$, the summed transmission of all channels, $\sum_{n=-N}^{N} \mathrm{T}_{k_{\shortparallel,n}}$, is supposed to be independent
of $k_{\shortparallel,\mathrm{max}}$. For this approach to be valid, a wide and short crystal must be
assumed, so that the details of the edges become less important.\cite{sepkhanov:063813}

\begin{figure}
\includegraphics[width=7.5cm]{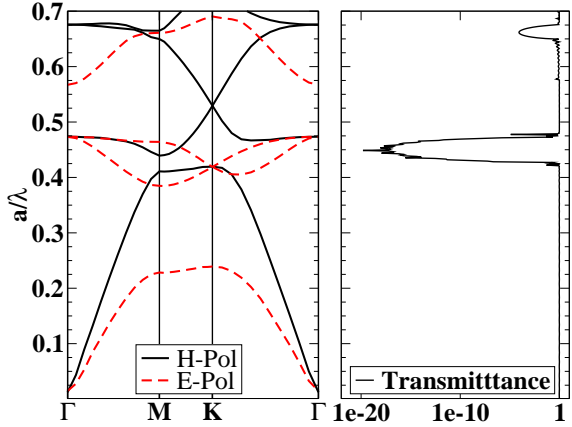}
\caption{\label{fig2}(Color online) Photonic band structure for dielectric cylinders in air on a hexagonal
lattice with $r/a=0.225$ and $\epsilon=14.0$. The arrow marks the Dirac point at $\omega'_D\,$=$\,0.5294$.
For H-polarization (magnetic field parallel to the cylinders) no propagating modes exist at
that frequency except for the one at the Dirac point. On the right side the transmittance for
H-polarization in the $\Gamma$K direction ($\theta\,$=$\,0^\circ$) is shown.}
\end{figure}

As a model system, we use cylinders ($r/a\,$=$\,0.225$, $\epsilon\,$=$\,14.0$) in air ($\epsilon\,$=$\,1.0$) on a hexagonal lattice. The crystal
orientation is chosen, so a plane wave at perpendicular incidence propagates along the $\Gamma$K direction, for which the Dirac
point occurs in H-polarization (magnetic field parallel to the cylinders). The corresponding band-structure,
together with the transmittance for H-polarization in the $\Gamma$K direction corresponding to
$\theta\,$=$\,0^\circ$, is shown in \myfigref{fig2}. The Dirac point
occurs at $\omega'_D\,$=$\,0.5294$ in the band-structure.

\begin{figure}\
\includegraphics[width=7.5cm]{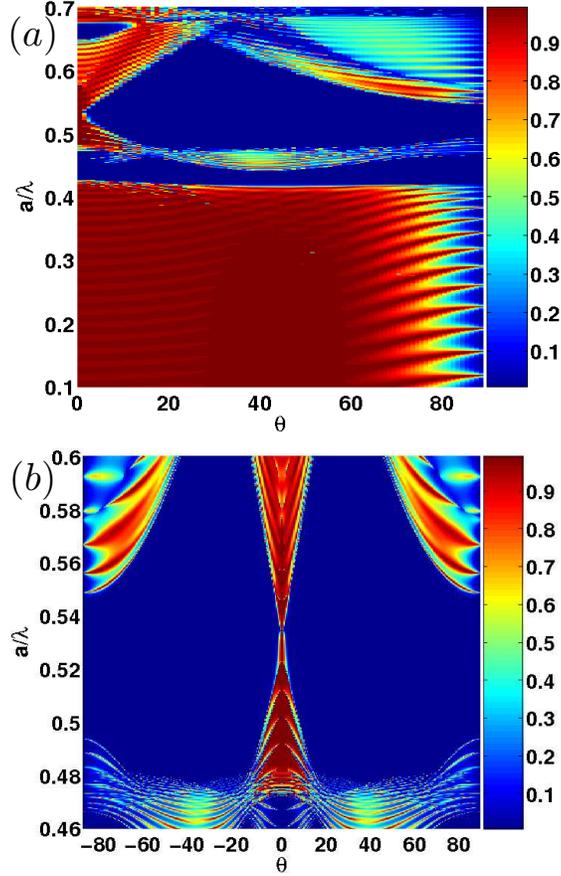}
\caption{\label{fig3}(Color online) (a) Angular and frequency-dependence of the transmittance (sum over all diffraction orders)
through a structure consisting of 40 cylinder rows for H-polarization. $\theta\,$=$\,0^\circ$ corresponds to
propagation in $\Gamma$K direction. The result is symmetric in the angle $\theta$.
(b) Enlargement around the Dirac point ($\omega'_D\approx0.532$).  The total transmittance, $\mathrm{T}$, discussed later in the paper, corresponds
to a summation of the transmittance over all angles with $\theta=\mathrm{arcsin}(nk_\shortparallel/k_0)$ $n=-N,\cdots, N$
(equidistant in $k_\shortparallel$ not in $\theta$) for a each frequency.}
\end{figure}

The dependence of the transmittance (sum over all propagating diffraction orders) on the angle and frequency of the
incident wave is shown in \myfigref{fig3} for a spacing of $\Delta\theta\,$=$\,1^{\circ}$.
The features of the band-structure, such as the stop band ($\omega'\approx0.45$), and the pseudo-stop band
around ($\omega'\approx0.675$) can be identified. \mbox{Figure\,\ref{fig3}b} enlarges the region around the Dirac point
for positive and negative angles. From the smallest width of the conical shaped transmittance, the Dirac point
can be estimated around $\omega'\approx0.532$.
However, resonances, due to the finite size, make a very precise determination more difficult and a better
way will be discussed later. In this case higher order diffraction orders do not
contribute significantly to the total transmittance. This is in contrast to the reflectance (not shown)
where for angles larger than 10--$15^{\circ}$ most energy is transferred in
the $\pm1^{\mathrm{st}}$ diffraction order, depending upon whether the angle
on the incoming wave is positive ($-1^{\mathrm{st}}$) or negative ($+1^{\mathrm{st}}$).
It should be noted, in these plots individual angles cannot be assigned a channel number, since the spacing is
not equidistant in $k_\shortparallel$.

Due to linear dispersion relation around the Dirac point, the phase of a plane wave with perpendicular incidence (in $\Gamma$K-direction)
changes linearly with frequency, if phase changes at the surfaces of the crystal are constant for all considered frequencies.
The phase change in the transmittance calculation, $\Delta \Phi_\mathrm{T}$, in a given frequency interval $\Delta\omega'$, is then equal
to the product of the length times the change of the wavevector in the band-structure in the same frequency interval
($\Delta \Phi_\mathrm{T}(\Delta\omega') = L\Delta k(\Delta\omega')$). Although not shown, our numerical results exhibit this
behavior extremely well. The absolute phase can only be determined up to an arbitrary but constant shift.
As will be shown later, a precise determination of the Dirac frequency is essential. However, since the band-structure
and transmittance are calculated by different methods, a small difference in the Dirac frequency
is found and the exact frequency for the transmittance calculations must be determined within the RCWA method.

\begin{figure}
\includegraphics[width=7.5cm]{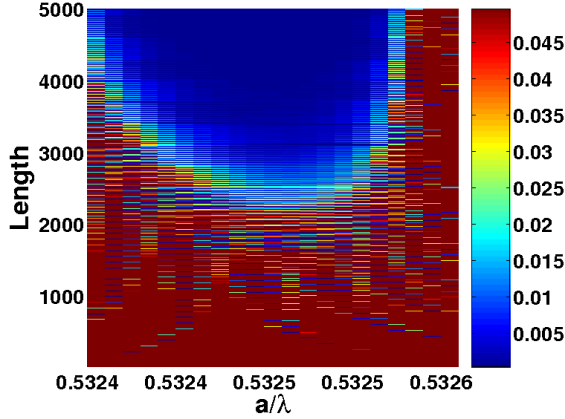}
\caption{\label{fig4}(Color online) Transmittance of the $1^{\mathrm{st}}$ channel
with $k_\shortparallel\,$=$\,2\pi/W\,$=$\,0.0013$ for
W=5000 over frequency. At the Dirac point  ($\omega'_D$=0.5325) the
transmittance is lowest for long structures.}
\end{figure}

This is possible by looking at the transmittance for a fixed $k_{\shortparallel}$, preferably close to zero,
and choosing the frequency for which this transmittance becomes the smallest for long structures. At the Dirac
point only the $k_{\shortparallel}=0.0$ component propagates in long structures. In \myfigref{fig4} the length-dependent
transmittance is plotted for different frequencies for $k_{\shortparallel}\,$=$\,0.00125$ corresponding to the $1^{\mathrm{st}}$
channel with a width of W\,=\,5000. We determine the frequency for the Dirac point to be $\omega'_D\,$=$\,0.5325$ for $\pm25$ modes
in the RCWA transmittance calculations. Although the transmittance typically converges better than $1\%$ with these numbers of modes, small shifts
in the frequencies still occur. Using only $\pm15$ modes instead of $\pm25$ changes the optimal value for the Dirac
frequency to 0.5318, corresponding to a shift of $0.13\%$. As a comparison,
the difference in the value from the band-structure (0.5294) corresponds to a difference of $0.58\%$. It is visible from \myfigref{fig4} that the frequency
must be determined precisely for large widths. A change in $\Delta\omega'$ of 0.0002 can turn the $1^{st}$ channel from non-propagating
to propagating, changing the scaling behavior significantly. In the band-structure, this would correspond to
going away from the Dirac point into the conical region, where a larger range of
$k_\shortparallel$ is available.

\begin{figure}
\includegraphics[width=7.5cm]{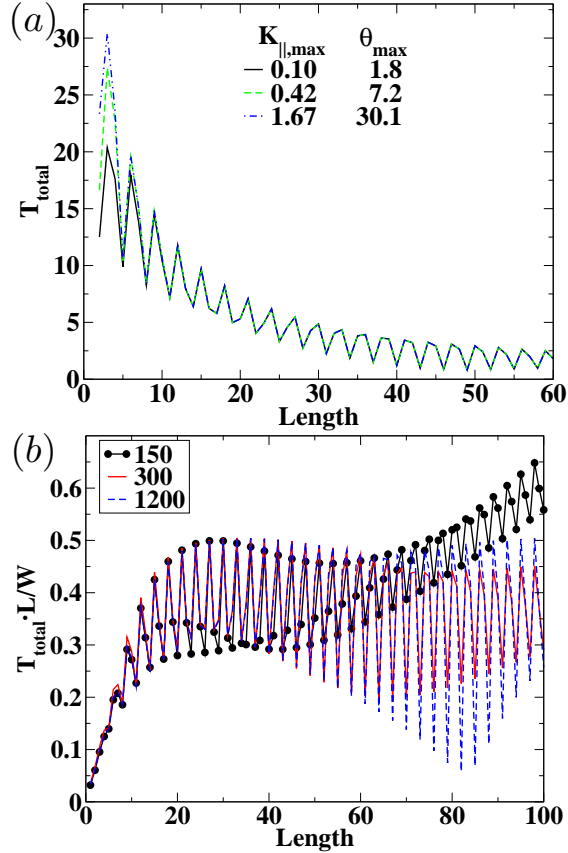}
\caption{\label{fig5}(Color online) (a) Total transmittance $\sum_{n=-N}^{N}\mathrm{T}_{k_{\shortparallel,n}}$ for
$N$=5, 20, and 80 with a fixed width of W=300 corresponding to different $k_{\shortparallel,\mathrm{max}}$ and
$\theta_{\mathrm{max}}$. The length, after which the results are equal in all cases, is determined by the smallest $k_{\shortparallel,\mathrm{max}}$.
(b) Normalized transmittance TL/W. The oscillations are caused by the finite size of the structure.
Different curves belong to different widths. The deviation of the black curve with the solid dots is due to the very narrow width. At approximately 40 unit cells,
only the channel with $k_{\shortparallel,\mathrm{max}}$=0 contributes with a constant transmittance. Hence the increase in the rescaled transmittance.}
\end{figure}

Using the determined Dirac frequency, we calculate the length- and width-dependence of the transmittance (\myfigref{fig5}a). For a fixed width of
$W\,$=$\,300$, we use a different number of channels corresponding to different $k_{\shortparallel,\mathrm{max}}$. Using more channels increases
the transmittance for short crystals, but, since channels with a large $k_{\shortparallel}$ decay rapidly, the total transmittance becomes independent
of this quantity after a length of approximately $1/k_{\shortparallel,\mathrm{max}}$.\cite{sepkhanov:063813}

According to the proposed scaling law, multiplying the total transmittance by a factor $\mathrm{L}/\mathrm{W}$ leads to a constant value.\cite{sepkhanov:063813}
Our results in \myfigref{fig5}b are not constant but oscillate around
a value of approximately 0.36, slightly higher than the predicted value of $1/\pi$.\cite{sepkhanov:063813} The oscillations 
in the transmittance, which depend on the length and surface termination of the crystal, are Fabry-Perot resonances caused by the finite length.
They do not exhibit a smooth curve since the sampling can only be completed in length steps of 1 unit cell and individual Fabry-Perot
oscillations are not resolved. They can be resolved by fixing the length and varying the frequency in very small steps.
Another deviation from the constant value can be seen in the curve for a width of 150 unit cells (curve with solid circles).
For this width, a linear increase in the normalized transmittance is visible, starting at a length of
40 unit cells. For structures longer than this width, only the $0^{\mathrm{th}}$ channel contributes to
transmittance with a constant value. The linear dependence of TL/W is then caused by the multiplication of T with the length of
the sample.

\begin{figure}
\includegraphics[width=7.5cm]{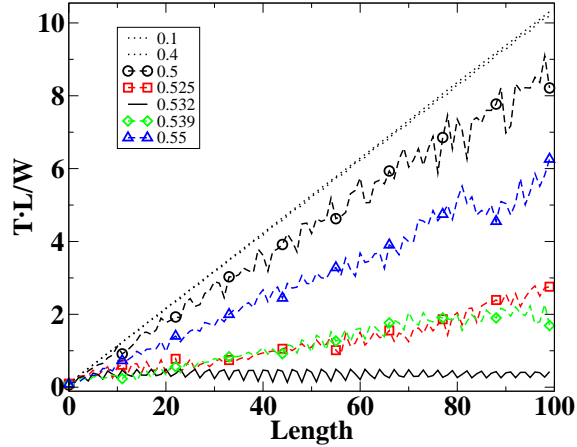}
\caption{\label{fig6}(Color online) Rescaled transmittance as a fuction of length 
for different frequencies ($N\,$=$\,10$, same $\theta_\mathrm{max}$ in all cases).
We can distinguish 3 regimes in the plot. Far ($\omega\,$=$\,0.1$) away from the Dirac point the rescaled transmittance
increases linearly. Here, all channels contribute with a high transmittance ($\approx 1.0$).
Close to the Dirac point the slope is reduced, since only a fraction of the channels
contributes. At the Dirac point ($\omega\,$=$\,0.532$) the curve oscillates around 0.36. The stop band region is not shown, but the
curve would be close to zero for all lengths due to the exponential decay.}
\end{figure}

\begin{figure}
\includegraphics[width=7.5cm]{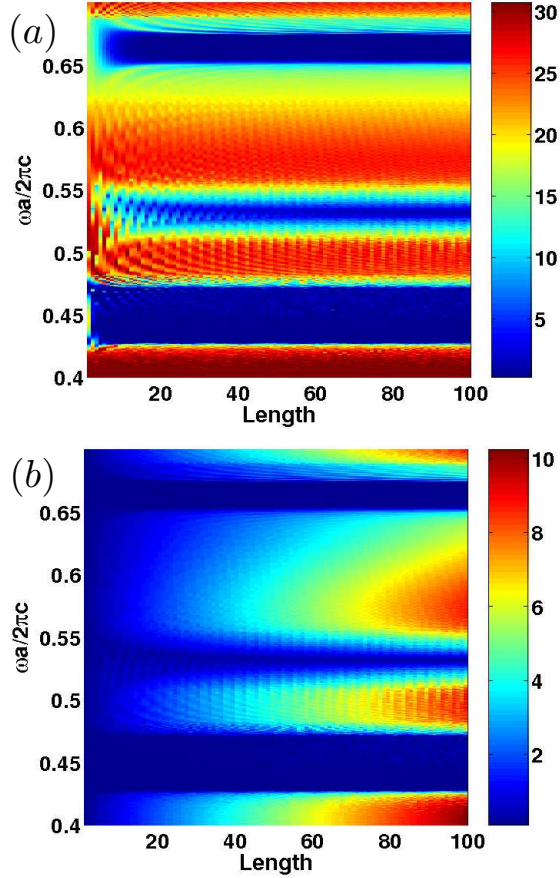}
\caption{\label{fig7}(Color online) (a): Total transmittance $\mathrm{T} = \sum_n \mathrm{T}_{k_{\shortparallel,n}}$ over length for
different frequencies (W=600,N=15). The three dark regions are the lower stop band
(0.45), the Dirac point (0.5325) and the quasi stop band (0.67). (b): Rescaled
transmittance TL/W. In the stop band this value is approximately zero;
whereas, at the Dirac point the value oscillates around 0.36.}
\end{figure}

\begin{figure}
\includegraphics[width=7.5cm]{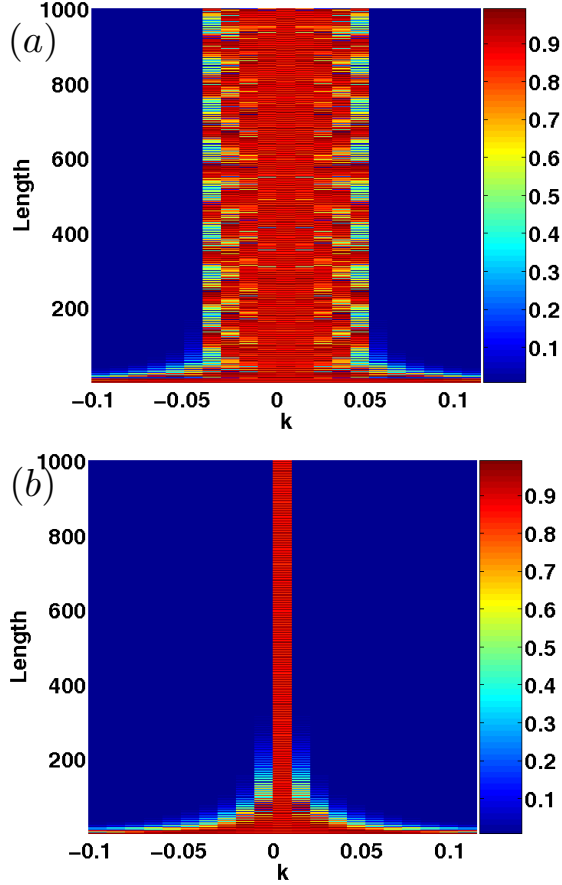}
\caption{\label{fig8}(Color online) Length-dependence of the first $\pm$10 channels for different frequencies.
(a) $\omega'\,$=$\,0.528$ corresponds to a line with an intermediate slope in \myfigref{fig6}.
(b) $\omega'\,$=$\,0.5325\,$=$\,\omega'_D$ belongs to the lowest line in \myfigref{fig6}. In both cases
the length scale is 10 times longer than in the previous graphs. The rescaled transmittance for
lengths after which the number of propagating channels stays constant (approx. 100 (250) in a (b))
exhibits the same behavior as the curve with solid dots in \myfigref{fig5}b at large lengths.}
\end{figure}

For different frequencies, we can identify several different characteristic behaviors for normalized transmittance (\myfigref{fig6}).
At low frequencies in the first band ($\omega\,$=$\,0.1$), normalized transmittance is given by a straight line with the same slope for all frequencies,
since all channels are contributing with a transmittance of about 100\%. The total transmittance $\mathrm{T}_\mathrm{total}$ then
corresponds to the number of channels.
In the second band, the transmittance still grows linearly with different slopes,
but oscillates around an average value.
In this case, some of the channels are contributing and the total number of contributing
channels determines the slope. At the Dirac point, a value around 0.36 is obtained
as discussed before. The final regime is the stop band (not shown in the plot), where
the normalized transmittance is always close to zero and decaying, since the effect
due to the exponential decay is stronger than the increase caused by multiplication
with the length. In all examples the maximum angle $\theta$ and the number
of channels are fixed. The four different regimes are also visible in
\myfigref{fig8}, which shows the total transmittance (a) and the normalized transmittance
(b) for a wide range of frequencies over length.

A better understanding for the occurrence of these 4 regimes can be obtained by looking at the length-dependent
transmittance of the individual channels for two example frequencies in \myfigref{fig8}.
Firstly, we consider a frequency close to the Dirac point (\myfigref{fig8}a). 
For short crystals up to about 100 unit cells, the number of propagating channels decreases and for longer
structures, nine channels contribute with a large transmittance. The rescaled transmittance is TL/W. Hence, it increases
linearly for a length exceeding 100 unit cells. At the Dirac point (\myfigref{fig8}b), all but the $0^{th}$ channel
are suppressed for long structures. In the region up to a length of about 250 unit cells, the suggested scaling
behavior is observed. Again, for longer structures TL/W will increase linearly, due to the constant transmittance
of the $0^\mathrm{th}$ channel comparable to the black curve with solid dots in \myfigref{fig5}b.

\begin{figure}
\includegraphics[width=7.5cm]{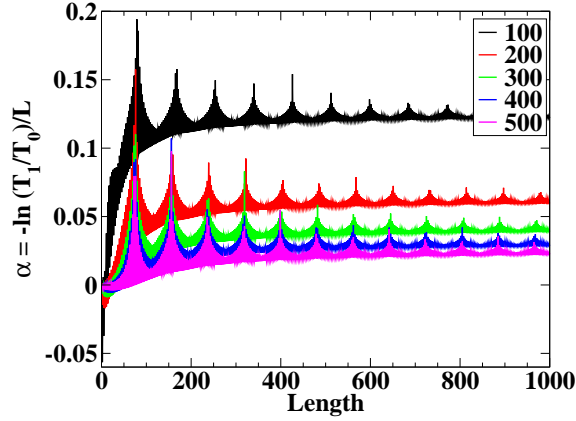}
\caption{\label{fig9}(Color online) Damping constant $\alpha = -\mathrm{ln}(\mathrm{T}_1/\mathrm{T}_0)/L$
of the $1^{st}$ channel with respect to the $0^{th}$ channel for different widths.
The highest curve corresponds to the structure with the smallest width.
The resonances are due to the finite length of the structure.}
\end{figure}

Previously, it has been stated that the scaling of the transmittance is valid for lengths larger than
$1/k_{\shortparallel,\mathrm{max}}$ in the limit of $W>>L$. As discussed before, there also exists an upper
limit for the length for this scaling behavior, which has not been addressed in previous publications.
We determined this
long length limit by comparing the $\pm1^\mathrm{st}$ channel to the $0^\mathrm{th}$ one. All channels with
$k_{\shortparallel}\neq0.0$ decay exponentially at the Dirac point, since there are no propagating
states available in the band-structure similar to the case in the gap. The propagating channels
always contribute with a transmittance of approximate unity,
so ignoring the details of the transmittance caused by the Fabry-Perot oscillations,
we can express the total transmittance of all channels by
\begin{equation}
\mathrm{T}_{\mathrm{total}} = \mathrm{T}_0 + \mathrm{T}_1 + \mathrm{T}_2 + \cdots = \mathrm{T} + \mathrm{e}^{-\alpha_1 L} \mathrm{T} + \mathrm{e}^{-\alpha_2 L} \mathrm{T} + \cdots
\end{equation}
with T on the order of 1. Since the $2^\mathrm{nd}$ and higher channels do not contribute significantly,
we can define the relative damping of the $1^\mathrm{st}$ channel with respect to the $0^\mathrm{th}$ as
$\alpha = -\mathrm{ln}(\mathrm{T}_1/\mathrm{T}_0)/L$, plotted for different widths in \myfigref{fig9}. As long as the
transmittance in the $1^\mathrm{st}$ channel decays, the damping constant increases until it saturates
and becomes constant.

\begin{figure}
\includegraphics[width=7.5cm]{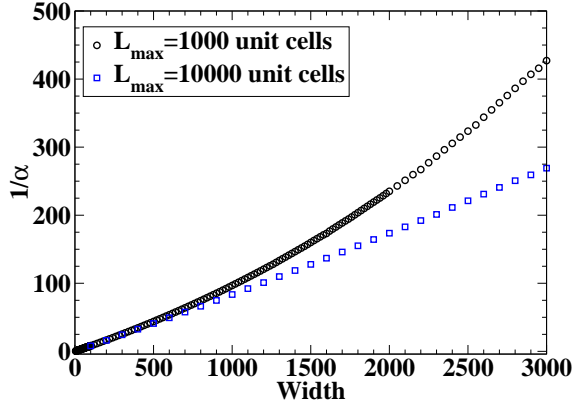}
\caption{\label{fig10}(Color online) Damping length $1/\alpha$ obtained by averaging
the damping constant on the left and inverting. If the crystal is not sufficiently long the damping constant does not reach
a constant value for very wide structures and the damping length deviates from
linear behavior (circles $L_{max}$=1000, squares $L_{max}$=10000).}
\end{figure}

To determine the maximum length for the 1/L scaling of the transmittance, the inverse
of the damping constant, given by the length for a suppression is $1/e$, can be used as a quantitative 
measure. This damping length is obtained by averaging the (length-dependent) damping constant, once it has reached
a constant value and then inverting the average. The averaging procedure is required to reduce
the impact of the Fabry-Perot resonances.

The results of this averaging are shown in \myfigref{fig10} and exhibit a linear behavior.
For crystals with a very wide width, the saturated value is only reached for very long lengths. Not
using sufficiently long structures leads to a damping constant, which is still increasing; hence, to an overestimation of the
damping length and a deviation from the linear behavior. From a linear fit, the slope can be obtained as 0.095,
which gives a width to length ratio of about 10:1, meaning the length limit of the W/L scaling of the transmittance
is reached at about 1/10 of the width. Identification of the upper limit is important, if one wants to study
disordered systems. In this case, the Dirac point may be shifted locally; hence, propagating modes are available
in regions where no modes were available before. Consequently, this can lead to an enhanced transmittance
in the vicinity of the Dirac point. If the channels with $k_{\shortparallel}\neq0.0$ are suppressed less compared
to $0^\mathrm{th}$ channel in the disordered case than in the unperturbed structure,
this will lead to a significant change in the damping length, even if
all channels experience changed due to disorder. Studying these quantities allows a better understanding of the
transmittance around the Dirac point in the case of disorder and offers the possibility to discuss the open question
whether disorder will increase or decrease the transmittance in this region.\cite{sepkhanov:063813}

In conclusion, we have presented detailed numerical calculations of the transmittance in hexagonal two-dimensional Photonic Crystals
close to the Dirac point. We found the transmittance at the Dirac point is inversely proportional
to the thickness of the sample. A detailed dependence of this behavior on the individual channels is given.
We give an explanation and a criterion for an upper length limit of this behavior and relate it to the width
of the crystal.
The dependence of the transmittance away from the Dirac point is also examined. It was determined that the transmittance
decays exponentially as expected when the frequency lies in the gap. When the frequency lies in the band,
not only the $k_{\shortparallel}\,$=$\,0$ component is contributing to the transmittance for all lengths.
The number of contributing channels depends on the width and the distance from the Dirac point frequency.

\begin{acknowledgments}
M.D. gratefully acknowledges financial support from the Alexander-von-Humboldt 
Foundation (Feodor-Lynen Program).
Work at Ames Laboratory was supported by the Department of Energy (Basic Energy Sciences)
under contract No. DE-AC02-07CH11358. This work was partially supported by the office
of Naval Research (Award No. N00014-07-1-0359)
\end{acknowledgments}


\end{document}